\documentclass[runningheads]{llncs}
\usepackage[T1]{fontenc}
\usepackage{graphicx}
\usepackage{xcolor}
\usepackage{tikz}
\usetikzlibrary{shapes.geometric, arrows.meta, positioning}

\usepackage{listings}

\newcommand\notodos{} 
\ifx\notodos\undefined
\usepackage{todonotes}
\else
\usepackage[disable]{todonotes}
\fi
\usepackage{csquotes} 
\usepackage[caption=false]{subfig}

\definecolor{pycomments}{rgb}{0.0, 0.5, 0.0}
\definecolor{pykeywords}{rgb}{0.0, 0.0, 0.8}
\definecolor{pystrings}{rgb}{0.6, 0.1, 0.1}
\definecolor{pybackground}{rgb}{0.98, 0.98, 0.98}
\definecolor{pynumbers}{rgb}{0.5, 0.5, 0.5}

\lstdefinestyle{pythonstyle}{
    language=Python,
    backgroundcolor=\color{pybackground},
    commentstyle=\color{pycomments}\itshape,
    keywordstyle=\color{pykeywords}\bfseries,
    stringstyle=\color{pystrings},
    numberstyle=\tiny\color{pynumbers},
    basicstyle=\ttfamily\footnotesize,
    breaklines=true,                   
    breakatwhitespace=false,           
    columns=flexible,                  
    float=htbp,                        
    captionpos=b,                      
    keepspaces=true,
    numbers=left,
    numbersep=8pt,
    showspaces=false,
    showstringspaces=false,
    showtabs=false,
    tabsize=4,
    frame=single,                      
    rulecolor=\color{pynumbers}
}

\def\packagename{BiJuTy}
\def\githubrepo{https://github.com/scads/bijuty}

\newcommand{\todoReply}[1]{\color{white}>~#1}

\usetikzlibrary{arrows.meta, positioning, fit, backgrounds, calc, shadows}

\usepackage{hyperref}

\begin{document}
%
\title{\packagename: An Interactive HPC-Aware Big Data Cluster Lifecycle Manager and Performance Assessment Utility for JupyterHub}
\titlerunning{\packagename}
%
\author{
Apurv Deepak Kulkarni\inst{1,2,3}\orcidID{0009-0003-3110-6801}%
\and
\\
Jan Frenzel\inst{2,3}\orcidID{0009-0007-5755-1427}%
\and
\\
Siavash Ghiasvand\inst{1,2,3}\orcidID{0000-0001-6627-0159}%
}
\authorrunning{Kulkarni et al.}
%
\institute{
Center for Scalable Data Analytics and Artificial Intelligence (ScaDS.AI) \and
Center for Interdisciplinary Digital Sciences (CIDS)\and
TUD Dresden University of Technology, Dresden, Germany\\
\email{\{%
apurv.kulkarni%
,%
jan.frenzel%
,%
siavash.ghiasvand%
\}@tu-dresden.de}
}
\maketitle              
\begin{abstract}
    The increasing demand for data processing has created a pressing need for access to high-performance computing (HPC) systems. Nevertheless, leveraging these systems to execute complex big data processing workflows remains a significant challenge, especially for beginners. This work presents \packagename{}, a solution designed to bridge the accessibility gap for big data workflows on HPC systems within the Jupyter ecosystem. By providing an interactive and user-friendly interface, \packagename{} simplifies cluster lifecycle management and performance assessment, making it more accessible on HPC systems to beginners and experienced users alike. The solution is presented as an interactive interface that guides the user through the entire process, from setting up the cluster configuration to carrying out initial performance assessments. Additionally, the framework enables seamless management of multiple clusters directly within the Jupyter Notebook interface, eliminating the need to switch outside of working environment. The collection of performance metrics from various sources further simplifies the optimization workflow. Furthermore, an illustrative example is provided to demonstrate how \packagename{} can be deployed to optimize the performance of a big data processing application. This example showcases how the entire big data processing lifecycle can be iteratively executed and optimized in just a few clicks, helping to reach the goal of optimization easily and interactively. By facilitating such workflows, this work contributes in bringing the field of big data computing and high-performance computing one step closer to the goal of seamless interaction and usability.
    \keywords{Big Data \and Jupyter \and HPC \and Slurm \and Apache Spark \and Apache Flink}
\end{abstract}
%
%
%
\section{Introduction}

The advancements in the field of machine learning and artificial intelligence (AI) increasingly motivate interdisciplinary researchers to use sophisticated analysis models.
This usage necessitates the access to larger computational resources such as high performance computing (HPC) clusters.
However, despite the wide-range of efforts and rapid advancements in democratizing HPC resources, there remains a significant gap which stems from the batch-processing nature of HPC clusters in contrast to the interactive approach of users who are coming from other fields.

Currently, the most accessible computation resources are the commercial cloud providers which provide HPC resources with a minimal barrier via their own graphical user interface (GUI)~\cite{microsoft2024managresources}.
However, despite the widespread availability and accessibility of cloud systems, HPC clusters remain a significant computation platform even for non-HPC workloads, due to reasons beyond user convenience. 
HPC clusters are purpose-built and have optimized architecture to provide extreme performance on specialized hardware within standardized environments while maintaining consistency across multiple runs. 
Additionally, increasing demands for data privacy and control motivate many institutions to maintain their own on-premise HPC clusters~\cite{agrawal2021,pramanik2021}.
These systems mainly run on open-source software and are tightly-coupled into the research environment over decades.

JupyterHub~\cite{jupyterhub} is one of the successful software infrastructures that bridges the gap and provides interactive access to HPC resources, particularly for users without prior experience, and is being actively used on both cloud and HPC clusters.
Its GUI requires only a modern web browser and serves as a single point of entry, while hiding all configurations from the user, which plays a significant role in the popularity of JupyterHub among the vast majority of users. %
Currently, many HPC centers provide JupyterHub as an alternative to the classical command line interface (CLI)-based access to their HPC clusters.
Although this interface cannot fully replace the CLI-based approach, since it does not align with the batch-processing nature of HPC systems, it provides an easy access to computational resources and assists the new users to get onboard and familiarize themselves with the HPC environment with minimal efforts.\todo{Maybe argue with pros and cons of GUI vs. CLI, e. g. trying things out vs. repeated experiments?\todoReply{I believe this would make introduction unnecessarily large. Maybe we can keep it is as last todo}}

Among the advanced analytical approaches, big data frameworks (BDF)~\cite{shahnawaz2025} are not exceptions.
BDFs are software environments and toolsets that are designed to efficiently ingest, store, process, and analyze massive volumes of data.
These frameworks have a wide range of usability in various interdisciplinary domains and often they need HPC resources to properly handle the vast amount of data.
However, setting up, managing, and monitoring BDF on HPC systems is itself a challenge for beginners due to a high level of complexity imposing a significant barrier on the path towards effective use of BDF on HPC systems. Beginners would significantly benefit from graphical representations of the configuration, guidance by predefined UI actions and simplified access to performance metrics.

Thus, this work proposes \packagename{}~\footnote{\githubrepo} a comprehensive infrastructure as code (IaC) toolbox that integrates into Jupyter ecosystem and extends its capabilities to (1)~manage individual users' configurations for various big data frameworks through a user-friendly interface, (2)~control the lifecycle of standalone big data clusters (BDC), (3)~visualize Spark/Flink job information, (4)~visualize framework and system-level metrics along with external metric systems, and (5)~manage configuration and workflow on a per-user and per-framework basis.

\packagename{} streamlines interaction with big data frameworks on HPC systems, making it accessible to novice users and efficient for experts.
It automates the set up, monitor, and tear down of BDCs.
Users will obtain performance-metric data with a single click, facilitating rapid debugging and deeper insight into underlying operations.
It enables users to manage configurations on an individual and per-framework basis.
All of these are provided directly from within the familiar JupyterHub ecosystem.

\section{Literature Review}
Several efforts focus on simplifying the access to HPC resources as a vital prerequisite for executing complex analysis pipelines which often need to handle large amount of data~\cite{chalker2021,samuel2021,liberati2025,sarker2024}.
Since this work is focusing on BDF, once an intuitive access to HPC resources is established e.g., via JupyterHub, the next step would be deployment of BDCs on the underlying HPC environment.
For deployment of BDCs on HPC clusters, many CLI tools are available for various HPC scheduling systems and either generic support for various BDC \cite{hanythingondemand,myhadoop,magpie} or tailored solutions for Apache Spark-based clusters \cite{spark-Slurm,spark-in-Slurm,sparkhpc,baer2015integrating}.
These tools reduce repetitive manual steps and streamline cluster setup for experienced practitioners.
However, this work focuses on automated mechanisms that minimize user intervention, thereby enabling researchers to concentrate on their core scientific objectives while abstracting the complexities of BDC management through systematic automations.

Various configuration management tools such as Chef~\cite{chef}, Puppet~\cite{puppet}, Ansible~\cite{ansible}, and Salt~\cite{salt} exist which can assist experienced users in installing and managing BDCs. However these tools are primarily designed for cloud environments with persistent infrastructure and full administrative access, rather than for HPC centers with resource scheduling and job-based execution environments. Consequently, they do not assist in provisioning BDCs within such environments and are targeted at experienced users.
Other automation tools such as Pulumi~\cite{pulumi} address the provisioning of underlying infrastructure, however most HPC clusters particularly in research institutes mainly rely on Slurm~\cite{slurm} or PBS~\cite{openpbs} as job scheduler which therefore requires a hybrid IaC approach.
Tools such as Open-OnDemand~\cite{open-ondemand} on the other hand have their specific hardware and software requirements and need deep integration into the software stack of the HPC clusters, which is out of scope for normal HPC users.
Therefore, although the above mentioned tools may simplify the deployment and provisioning of BDCs, none of them can provide an accessible approach for beginners.

Besides community efforts such as BDWatchdog~\cite{enes2018bdwatchdog}, that provides an accessible approach for monitoring and profiling BDF, many attempts have also been made by maintainers of BDF such as Spark~\cite{apache_spark_web_ui} and Flink~\cite{apache_flink_webui_playground} to make monitoring of these frameworks more accessible and user-friendly.

Another approach, that is particularly beneficial, is maintaining JupyterHub as the central entry point while integrating the relevant features of BDCs directly into Jupyter ecosystem.
With this approach, users can access the required functionality directly from within the familiar JupyterHub interface.
Streaming Jupyter integration~\cite{streamingJupyterIntegration}, as an example, enables interactive execution of Flink SQL jobs in Jupyter notebooks, SparkMonitor~\cite{swan_cern_sparkmonitor} embeds the resource monitoring of Spark resources into Jupyter notebooks, and JupyterLab Spark~\cite{jupyterlabSpark} integrates the native Spark web UI into the JupyterHub interface.
Although these approaches contribute to the accessibility of BDCs monitoring, to the best of our knowledge, no tool exists that combines configuration, provisioning and monitoring of BDC within JupyterHub interface.

\section{\packagename}

\packagename{}, as described before, is an Interactive HPC-Aware BDC lifecycle manager and performance assessment utility that integrates its features seamlessly into JupyterHub providing a consistent and flexible environment.
Current implementation is compatible with all HPC clusters that utilize Slurm batch job scheduler.

\subsection{Workflow}
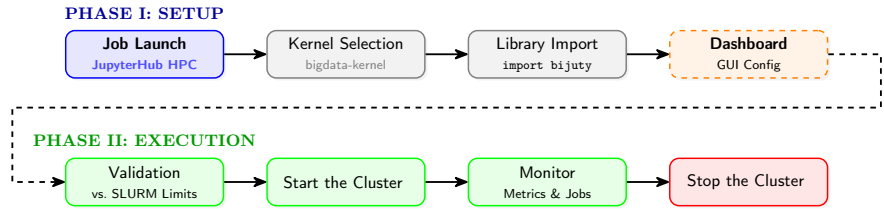
\begin{figure}[ht]
    \centering
    \scalebox{0.95}{\resizebox{\textwidth}{!}{%
        \begin{tikzpicture}[
    node distance = 0.5cm and 0.8cm,
    every node/.style = {font=\sffamily\small},
    box/.style = {rectangle, draw=gray!20, fill=white, minimum width=3cm, minimum height=9mm, rounded corners=5pt, align=center, line width=0.5pt, drop shadow={opacity=0.1, shadow xshift=0.5mm, shadow yshift=-0.5mm}, thick},
    colorstyle/.style = {fill=#1!10, draw=#1},
    start_style/.style = {box, colorstyle=blue, font=\sffamily\bfseries},
    proc_style/.style = {box, colorstyle=gray},
    mode_style/.style = {box, colorstyle=orange, dashed},
    deploy_style/.style = {box, colorstyle=green},
    stop_style/.style = {box, colorstyle=red},
    arrow/.style = {-{Stealth[round, scale=1.2]}, line width=1pt, draw},
]

    \node [start_style] (launch) {Job Launch \\ \scriptsize \color{blue!70} JupyterHub HPC};
    \node [proc_style, right=of launch] (kernel) {Kernel Selection \\ \scriptsize \color{gray} bigdata-kernel};
    \node [proc_style, right=of kernel] (import) {Library Import \\ \scriptsize \texttt{import \MakeLowercase{\packagename{}}}};

    \node [mode_style, right=of import] (beginner) {\textbf{Dashboard} \\ \scriptsize GUI Config};

    \node [deploy_style, below=15mm of launch] (validate) {Validation \\ \scriptsize vs. SLURM Limits};
    \node [deploy_style, right=of validate] (start) {Start the Cluster
    };
    \node [deploy_style, right=of start] (monitor) {Monitor \\ \scriptsize Metrics \& Jobs};
    \node [stop_style, right=of monitor] (stop) {Stop the Cluster
    };

    \draw [arrow] (launch) -- (kernel);
    \draw [arrow] (kernel) -- (import);
    
    \draw [arrow] (import.east) .. controls +(0.4,0) and +(-0.4,0) .. (beginner.west);
    
    
    \node (center) [yshift=2mm] at ($(launch)!0.5!(validate)$) {};
    \draw[arrow, dashed, draw]
  (beginner.east)
  -- ++(10mm,0mm)
  -- ([xshift=10mm] beginner.east |- center)
  -- ([xshift=-10mm] validate.west|- center)
  -- ([xshift=-10mm] validate.west)
  -- (validate.west);

    \draw [arrow] (validate) -- (start);
    \draw [arrow] (start) -- (monitor);
    \draw [arrow] (monitor) -- (stop);

    \node[above=0.1cm of launch, font=\bfseries\color{blue!60!black}] {PHASE I: SETUP};
    \node[above=0.1cm of validate, font=\bfseries\color{green!60!black}] {PHASE II: EXECUTION};

\end{tikzpicture}%
    }}%
    
    \caption{\packagename{} schematic workflow}
    \label{fig:jubility-init-workflow}
\end{figure}

The workflow, as illustrated in \figurename~\ref{fig:jubility-init-workflow}, starts by initiation of JupyterHub as a Slurm job on HPC cluster.
Then, an appropriate pre-configured kernel (e.g. 'bigdata-kernel') should be selected which in turn, opens a notebook.
In the notebook, the user imports the \packagename{} library in one of the cells (ideally at the top).
Upon completion of the import, the user is presented with a interface that displays all existing settings.
Subsequently, the user can proceed to the next stages of the workflow.

\begin{figure}[ht]
    \tikzset{
      mylabel/.style = {draw, rounded corners, text width=20mm, align=center,}
    }
    \centering
    \scalebox{0.98}{
    \begin{tikzpicture}
        \node(main){\includegraphics[width=\textwidth]{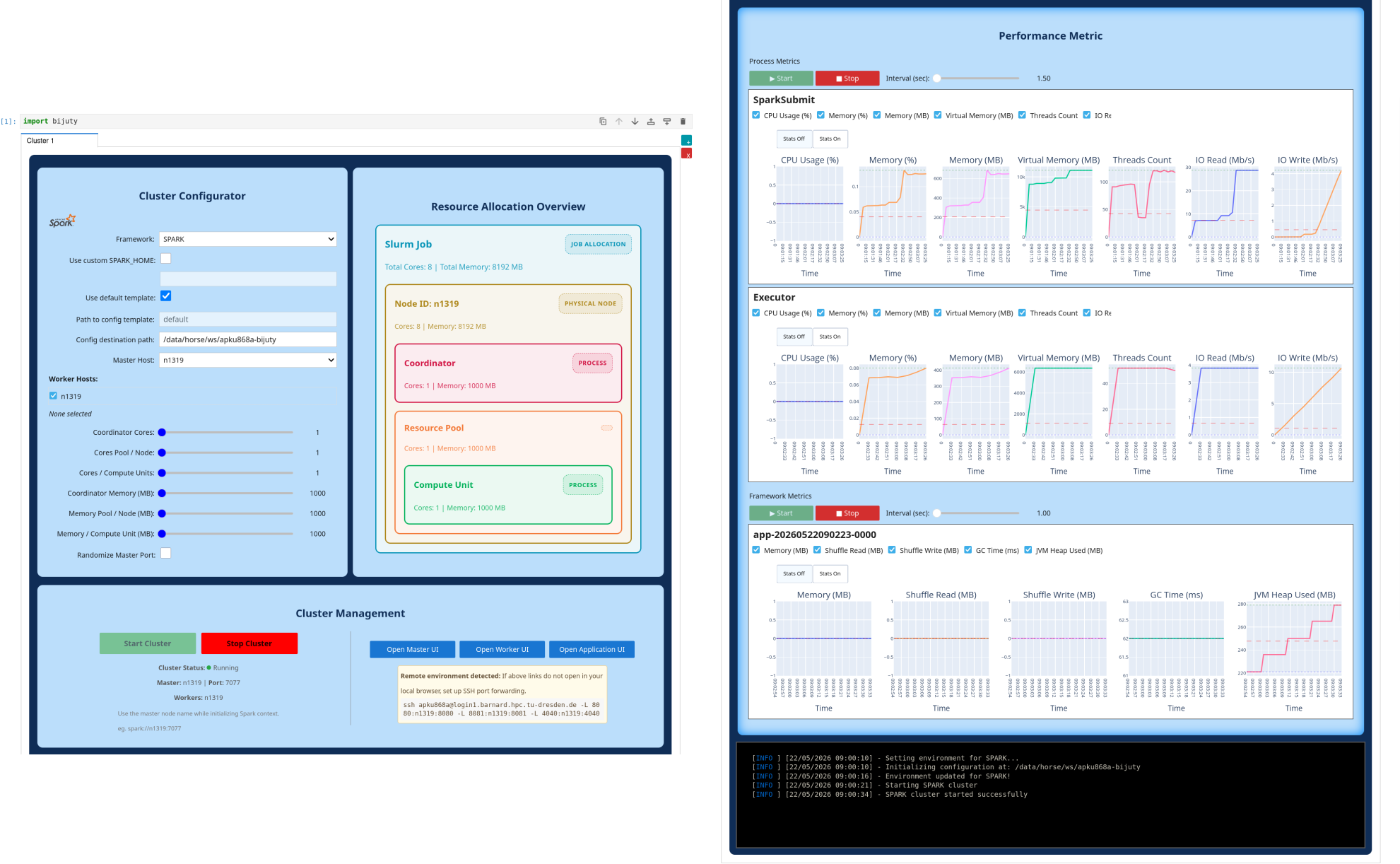}};
        \draw[-,draw=red, very thick] (main.south)++(-62mm,10.25mm) -- ++(28mm,0) -- ++(2mm,2mm) -- ++(2mm,-4mm)-- ++(2mm,2mm) -- ++(28mm,0mm);
        \draw[-,draw=red, very thick] (main.north)++(0mm,-0.5mm) -- ++(28mm,0mm) -- ++(2mm,2mm) -- ++(2mm,-4mm)-- ++(2mm,2mm) -- ++(28mm,0mm);
    \end{tikzpicture}
    }
    \caption{\packagename{} interface}
    \label{fig:jubility-interface}
\end{figure}

The interface presents different options and buttons to manage the standalone big data cluster, as shown in \figurename~\ref{fig:jubility-interface}.
It consists of different sections such as \textquote{Cluster Configurator}, \textquote{Resource Allocation Overview}, \textquote{Cluster Controls}, \textquote{Performance Metric} and interface log panel.
Sections are (de-)activated accordingly to guide the user through out the process.
Tooltips are also offered where additional information or context is required.

The \textquote{Cluster Configurator} section consists of basic options required to setup a standalone cluster.
The default values are set such that they are sufficient for performing basic computation for testing purpose.
Parameters in this section utilize generic names to ensure adaptability to various naming conventions used in major BDFs.
For example, Apache Spark uses the notion of a \emph{Driver}~\cite{sparkcluster} for the component which steers distributed execution while such a component is named \emph{JobManager}~\cite{flinkcluster} in case of Apache Flink.
The interface uses the term \emph{Coordinator} instead to emphasize its role.
User may initialize the standalone cluster using the provided default values or can use their own configuration template for a more customized usage, with an optional unique path for setting location of initialization. %
%
The section \textquote{Resource Allocation Overview} provides a high-level visual overview about resource distribution among different processes and nodes inside Slurm job. The visualization projects a dynamic overview according to the user's configuration. %
A standalone cluster can be started and stopped from the Cluster Control section.
Once the required parameters are set, the configuration can be loaded to the current environment and a standalone cluster can be started by clicking on \textquote{Start Cluster} button.
In current implementation, only standalone cluster is supported for Apache Spark and Apache Flink frameworks. The integrated web interfaces provided by each BDF can be also accessed using the relevant buttons directly from \packagename{} interface. %
Following a successful start of the cluster processes, the \textquote{Performance Metric} section is enabled for interaction, providing a performance overview. All desired processes', framework's and external metrics can be monitored in real-time using the Performance Metrics section.

These visualizations and overviews provide users with a deeper understanding of big data processing workflow. 
\packagename{} allows users to execute the workflow iteratively to experiment with various configurations.
Furthermore, multiple frameworks such as Apache Flink and Apache Spark can be used and managed simultaneously within a single interface.
Users can manage these views by adding or removing tabs via the buttons located in the top-right corner.

\begin{figure}[ht]
    \centering
    \resizebox{\textwidth}{!}{%
        \begin{tikzpicture}[
    node distance = 0.5cm and 0.8cm,
    every node/.style = {font=\sffamily\small},
    box/.style = {rectangle, draw=gray!20, fill=white, minimum width=3cm, minimum height=1.1cm, rounded corners=5pt, align=center, line width=0.5pt, drop shadow={opacity=0.1, shadow xshift=0.5mm, shadow yshift=-0.5mm}},
    colorstyle/.style = {fill=#1!10, draw=#1},
    node_style/.style = {box, colorstyle=blue, font=\sffamily\bfseries},
    sched_node_style/.style = {box, colorstyle=green, font=\sffamily\bfseries},
    proc_style/.style = {box, colorstyle=gray},
    bijuty_session_style/.style = {box, colorstyle=orange, font=\sffamily},
    arrow/.style = {-{Stealth[round, scale=1.2]}, line width=1pt, draw=gray},
    slurmarrow/.style = {-{Stealth[round, scale=1.2]}, line width=1pt, draw=green},
    bdcarrow/.style = {-{Stealth[round, scale=1.2]}, line width=1pt, draw=gray},
    jupyterhubarrow/.style = {-{Stealth[round, scale=1.2]}, line width=1pt, draw=orange},
]

    \node [node_style] (slurm_scheduler) {Slurm scheduler \\ \scriptsize \color{blue!70} Available Infrastructure};
    \node [node_style, right=of slurm_scheduler] (slurm_1) {Compute node 1 \\ \scriptsize \color{blue!70} Available Infrastructure};
    \node [sched_node_style, right=of slurm_1] (slurm_master) {Compute node 2 \\ \scriptsize part of Slurm job};
    \node [sched_node_style, right=of slurm_master] (slurm_worker1) {Compute node 3 \\ \scriptsize part of Slurm job};
    \node [node_style, right=of slurm_worker1] (slurm_4) {Compute node 4 \\ \scriptsize \color{blue!70} Available Infrastructure};
    \node [sched_node_style, right=of slurm_4] (slurm_worker2) {Compute node 5 \\ \scriptsize part of Slurm job};

    \node [node_style, above=of slurm_scheduler] (jupyterhub_spawner) {Jupyterhub Spawner \\ \scriptsize \color{blue!70} Available Infrastructure};
    
    \node [bijuty_session_style, above=of slurm_master] (bijuty_session) {\packagename{} session \\ \scriptsize \color{gray} runs user notebook};
    \node [proc_style, above=of bijuty_session] (bdc_master) {BDC Master \\ \scriptsize \color{gray} schedules BDC tasks};
    \node [proc_style, above=of slurm_worker1] (bdc_worker1) {BDC Worker1 \\ \scriptsize \color{gray} runs BDC tasks};
    \node [proc_style, above=of slurm_worker2] (bdc_worker2) {BDC Worker2 \\ \scriptsize \color{gray} runs BDC tasks};

    \draw [jupyterhubarrow] (jupyterhub_spawner) -- (slurm_scheduler);
    

    \node(left_anchor)[below=5mm of slurm_1.south west] {};
    \draw [slurmarrow] (slurm_scheduler.south) -- (slurm_scheduler.south east|-left_anchor.center) -- (slurm_master.south west|-left_anchor.center) -- (slurm_master.south);
    
    \draw [slurmarrow] (slurm_scheduler.south) -- (slurm_scheduler.south east|-left_anchor.center) -- (left_anchor.center-|slurm_worker1.south west) --  (slurm_worker1.south);
    
    \draw [slurmarrow] (slurm_scheduler.south) -- (slurm_scheduler.south east|-left_anchor.center) -- (left_anchor.center-|slurm_worker2.south west) -- (slurm_worker2.south);
    
    \draw [jupyterhubarrow] (jupyterhub_spawner) -- (bijuty_session);
    
    \draw [bdcarrow] (bijuty_session) -- (bdc_master);
    \draw [bdcarrow] (bijuty_session) -- (bdc_worker1);
    
    \node(top_left_anchor)[above=5mm of bdc_worker1.north west] {};
    \node(top_right_anchor)[at=(slurm_4.north east|-top_left_anchor)] {};
        
    \draw [bdcarrow] (bijuty_session.north east) -- 
            (top_left_anchor.center) -- (top_right_anchor.center) --
            (bdc_worker2.north west);
    

\end{tikzpicture}
    }

    \caption{Interactions between JupyterHub, Slurm, BiJuTy and the BDC.}
    \label{fig:workflow-birdseye}
\end{figure}
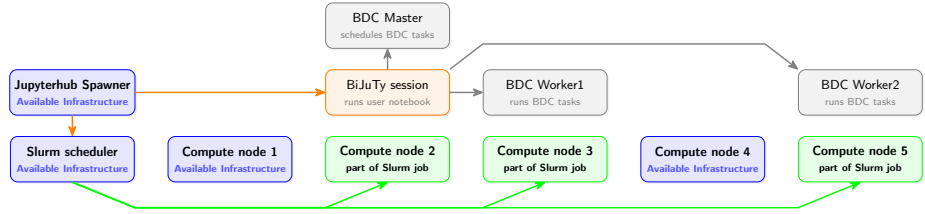

\figurename~\ref{fig:workflow-birdseye} further extends the schematic workflow illustrated in \figurename~\ref{fig:jubility-init-workflow} by incorporating the interactions between the JupyterHub, Slurm and BDC. The JupyterHub Spawner instructs Slurm to schedule a job. On one of the allocated resources (green boxes), the JupyterHub Spawner starts a Jupyter process, in which a \packagename{} session is started. Via the \packagename{} session (orange), a user can configure and start the processes for the BDC (gray boxes).

\subsection{Implementation}

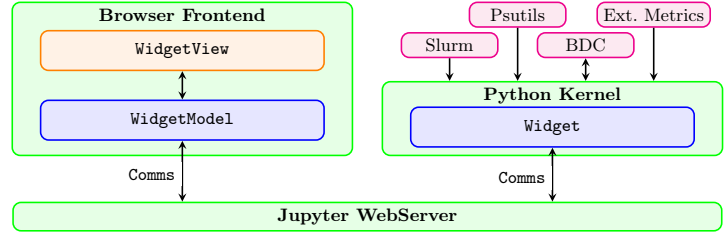
\begin{figure}[ht]
    \centering
    \scalebox{0.75}{\begin{tikzpicture}[
    box/.style = {rounded corners, fill=green!10, draw=green, font=\bfseries, thick},
    boxp/.style = {rounded corners, fill=#1!10, draw=#1, anchor=center, thick},
    wh/.style 2 args = {minimum width=#1,minimum height=#2}
]
    \node(frontend)[box,wh={60mm}{27mm}]{};
    \node(frontend-label)[box,fill=none,draw=none, below=0mm of frontend.north]{Browser Frontend};
    \node(widgetView)[boxp=orange, wh={50mm}{7mm}, below=5mm of frontend.north]{\texttt{WidgetView}};
    \node(widgetModel)[boxp=blue, wh={50mm}{7mm}, below=5mm of widgetView]{\texttt{WidgetModel}};
    \node(backend)[box,wh={60mm}{13mm}, right=5mm of frontend.south east, anchor=south west]{};
    \node(backend-label)[box,fill=none,draw=none, below=0mm of backend.north]{Python Kernel};
    \node(widget)[boxp=blue, wh={50mm}{7mm}, at=(backend.center), yshift=-1.5mm]{\texttt{Widget}};
    \node(slurm)[boxp=magenta, wh={17mm}{4mm}, above=4mm of backend.north west, xshift=12mm]{Slurm};
    \node(psutil)[boxp=magenta, wh={17mm}{4mm}, at=(slurm.center|-frontend.north), anchor=north, xshift=12mm]{Psutils};
    \node(bdc)[boxp=magenta, wh={17mm}{4mm}, at=(psutil.center|-slurm.north), anchor=north, xshift=12mm]{BDC};
    \node(exm)[boxp=magenta, wh={17mm}{4mm}, at=(bdc.center|-psutil.north), anchor=north, xshift=12mm]{Ext. Metrics};
    \node(center)[at=($(frontend)!0.5!(backend)$)]{};
    \node(webserver)[box, minimum height=5mm,minimum width=125mm, below=17mm of center]{};
    \node(webserver-label)[box,,fill=none,draw=none, at=(webserver.center)]{Jupyter WebServer};
    \draw[stealth-stealth, thick, black] (widgetView) -- (widgetModel);
    \draw[stealth-stealth, thick, black] (widgetModel) --node[midway, left,fill=white, yshift=-0.5mm]{\texttt{Comms}} (widgetModel|-webserver.north);
    \draw[stealth-stealth, thick, black] (widget|-webserver.north) --node[midway,left, fill=white, yshift=-0.5mm]{\texttt{Comms}} (widget);
    \draw[-stealth, thick, black] (slurm) -- (slurm|-backend.north);
    \draw[-stealth, thick, black] (psutil) -- (psutil|-backend.north);
    \draw[stealth-stealth, thick, black] (bdc) -- (bdc|-backend.north);
    \draw[-stealth, thick, black] (exm) -- (exm|-backend.north);

\end{tikzpicture}}
    \caption{\packagename{} integration in JupyterHub (adapted from~\cite{ipywidget}).}
    \label{fig:widget-arch}
\end{figure}

The \packagename{} package follows Jupyter's widget architecture, as shown in \figurename~\ref{fig:widget-arch}.
This architecture facilitates maintaining synchronized widget state between the Python kernel and the frontend, thus helping in decoupling python backend from GUI.
The client side GUI is implemented using IPyWidgets~\cite{ipywidget}.
The widget that is loaded in the kernel interfaces with backend components such as Slurm, \texttt{psutil}, external metric collectors and BDF backend.
The widgets used in the \packagename{} package have representations both in Python kernel (\texttt{Widget}) and the frontend (\texttt{WidgetModel}).
A bidirectional state synchronization exists between these objects through Jupyter's communication protocol (\texttt{Comms})~\cite{jupytercomm}.
Displaying a widget creates a \texttt{WidgetView} instance representing that \texttt{WidgetModel}.
Any update from the \texttt{WidgetView} goes to the \texttt{WidgetModel} and synchronizes with the \texttt{Widget}.
Similarly, Python kernel outputs, such as data processing output or plot, are sent to the \texttt{Widget} which then synchronizes with the \texttt{WidgetModel} to update the \texttt{WidgetView}.

The Slurm interface present in this package extracts the job information using Slurm's native commands and the environment variables. In Slurm-based HPC environments, software packages are typically installed centrally on shared filesystems and made available cluster-wide through environment modules or software stacks. Since these frameworks may be accessed concurrently by multiple jobs and users, it is crucial to ensure that the individual user configuration remain isolated and do not interfere with each other. To address this requirement, \packagename{} uses a template approach from \texttt{BigDataFrameworkConfigure}~\cite{bigdataframeworkconfigure}, where the Python kernel derives a configuration from framework-specific templates and values that are present in the interface, and initializes an isolated configuration along with setting environment variables required by BDFs.

The Python kernel also uses the configuration to invokes bash scripts of the selected BDF, in order to start or stop the BDC.
The metrics are collected asynchronously across three different levels, namely: Process, Framework and External levels.
The Process level metrics are collected using \texttt{psutil}~\cite{psutil} Python package, that collects metrics such as CPU and Memory usage.
Additionally, metrics such as heap memory usage, garbage collection (GC) statistics, and other relevant metrics are collected using the API endpoints provided by BDFs. 

The metrics on External level are collected from external services such as \emph{Pika}~\cite{Dietrich2020-jx}.
Pika in this case provides node level metrics including Power usage and IO Read/Write among the others.
All these metrics are plotted and regularly updated at an interval that displays system status near real-time.


\section{Results and discussion}

This section describes a workflow for deploying the \packagename{} package and utilizing it for the use-case of assessing performance of and application through an experimental evaluation.

\begin{figure}
    \centering
    \scalebox{0.75}{
            

\begin{tikzpicture}[
    boxcolor/.style={draw=#1, fill=#1!10},
    boxdim/.style 2 args={minimum width=#1, minimum height=#2},
    startstop/.style={rectangle, rounded corners, boxcolor=#1, boxdim={20mm}{10mm}},
    process/.style={rectangle, boxcolor=blue,  align=center,boxdim={25mm}{10mm}},
    decision/.style={diamond, draw, boxcolor=orange, aspect=2, align=center},
    myarrow/.style={-{Stealth[length=3mm,width=4mm]}, line width=1mm, draw=blue},
    dashboard/.style={dashed, very thick}
]

\def\radius{25mm}

\node [startstop=green, xshift=-50mm] (begin) at (90:\radius) {Start};
\node [startstop=red, xshift=-50mm] (end) at (-90:\radius) {Finish};

\node [process,dashboard] (start)   at (90:\radius) {Start cluster};
\node [process] (submit)  at (30:\radius)   {Submit\\application};
\node [process,dashboard] (observe) at (-30:\radius) {Observe\\performance};
\node [decision] (decide) at (-90:\radius) {Satisfactory?};
\node [process,dashboard] (stop) at (-150:\radius) {Stop cluster};
\node [process,dashboard] (revise)  at (-210:\radius) {Revise\\configuration};

\node(legend)[overlay, remember picture, process,dashboard, right=42mm of decide.center, minimum width=5mm, minimum height=5mm] {};
\node(legend-text)[overlay, remember picture, right=2mm of legend, text width=15mm] {Interaction in \packagename{} interface};

\draw[myarrow, draw=green] (begin) -- (start);
\def\bend{15}
\draw[myarrow] (start)   to[bend left=\bend] (submit);
\draw[myarrow] (submit) to[bend left=\bend] (observe);
\draw[myarrow] (observe) to[bend left=\bend] (decide);
\draw[myarrow] (decide) to[bend left=\bend]node[midway,above, shift=({3mm,-1.5mm})]{No} (stop);
\draw[myarrow] (stop) to[bend left=\bend] (revise);
\draw[myarrow] (revise) to[bend left=\bend] (start);
\draw[myarrow, draw=red] (decide) -- node[midway,above]{Yes} (end);
\end{tikzpicture}



}
    \caption{Optimization use-case workflow using \packagename{}.
    }
    \label{fig:use-case-workflow}
\end{figure}
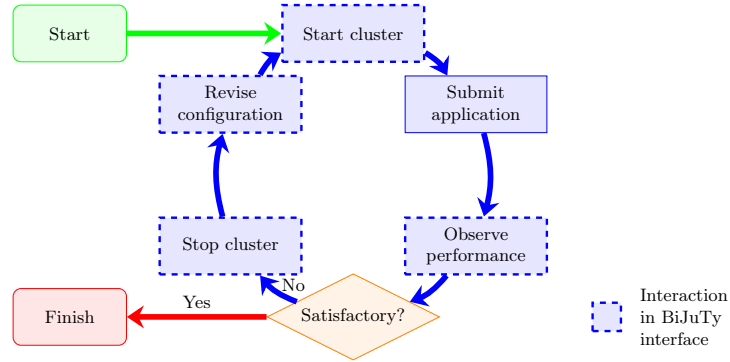

The experiment follows the optimization cycle illustrated in \figurename~\ref{fig:use-case-workflow}. Initially, the standalone cluster is started from the \packagename{} interface and the application is submitted with a baseline configuration. Application performance is then observed. If the observed performance is unsatisfactory, the cluster is stopped and restarted from the interface with a revised configuration; the application is subsequently resubmitted to the reconfigured cluster, and performance metrics are reassessed. For this demonstration, the cycle is limited to a single iteration.

\packagename{}'s applicability is demonstrated on the Barnard system of the ZIH-Clusters~\cite{tudresdenHPCSysteme}, which uses Slurm for batch scheduling and provides shared access to Apache Spark installation via Module System~\cite{modulesystem}. A JupyterHub session is launched and attached to a Slurm job with 8 CPUs and 8GB of total memory on single node. Upon successful spawning of JupyterHub, the \packagename{} package can be installed by running the following command inside a notebook cell: \texttt{pip install git+\githubrepo@main}. This step needs to be done only once for any Python kernel that is being used. Following the installation, the web browser page is reloaded and the Python kernel is restarted, after which, this kernel is used for all subsequent operations.

Apache Spark's PySpark~\cite{pyspark} is used to showcase how \packagename{} can help in running the complete big data processing workflow and assessing the impact of configuration changes on performance metrics. The workload~\footnote{\githubrepo/tree/main/example} comprises of a data processing and analysis on Uber taxi pickup data~\cite{ubertlcfoilresponse} comprisng 4+ Million  rows of location, date and time and base codes of compnies affiliated to Uber pickup. The example involves grouping and aggregation computation.


\begin{figure}[ht]
    \centering
    \subfloat[Phase I\label{fig:test-setup-phaseI}]{\includegraphics[width=0.40\textwidth]{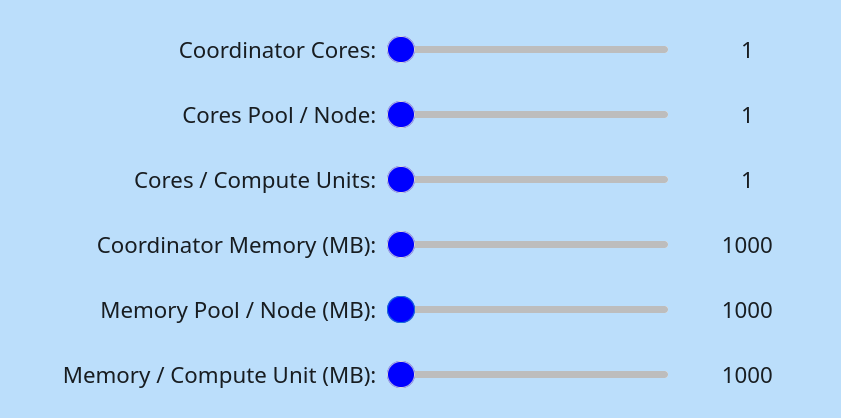}}        
    \hspace{5mm}
    \subfloat[Phase II\label{fig:test-setup-phaseII}]{\includegraphics[width=0.40\textwidth]{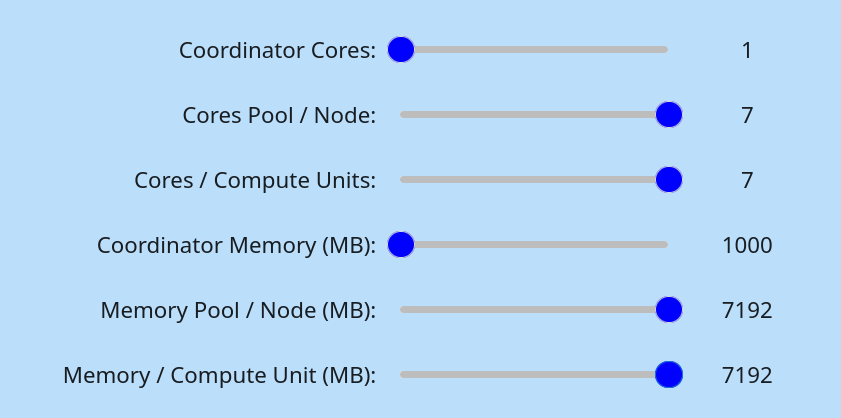}}        
    \caption{Cluster Parameters}
    \label{fig:test-setup}
\end{figure}

The \packagename{} interface is launched by importing the \texttt{\MakeLowercase{\packagename{}}} package (\texttt{import \MakeLowercase{\packagename{}}}). In this interface, Apache Spark is chosen as a framework of choice for the experiment. The evaluation is conducted in two phases, one with lower resources and one with higher resources with the aim of checking if the higher resource really improves the performance or not. For phase I, parameters are set to minimum before deploying Apache Spark standalone cluster, as illustrated in \figurename~\ref{fig:test-setup-phaseI}. Upon successful startup of the cluster, the section \textquote{Performance Metric} activates, allowing direct access to performance data on different levels, as discussed before. Phase II of the experiment can be executed with increased resources by stopping the cluster, restarting it with revised parameter values (as presented in \figurename~\ref{fig:test-setup-phaseII}), and re-running the same PySpark code. It can be observed that the workload executes with increased memory allocation and CPUs count within each compute unit (\emph{Executor} in Apache Spark~\cite{sparkcluster}).

\begin{figure}[ht]
    \centering
    \scalebox{0.96}{\begin{tikzpicture}
        \node(main)[minimum width=\textwidth, minimum height=30mm]{\includegraphics[width=0.95\textwidth]{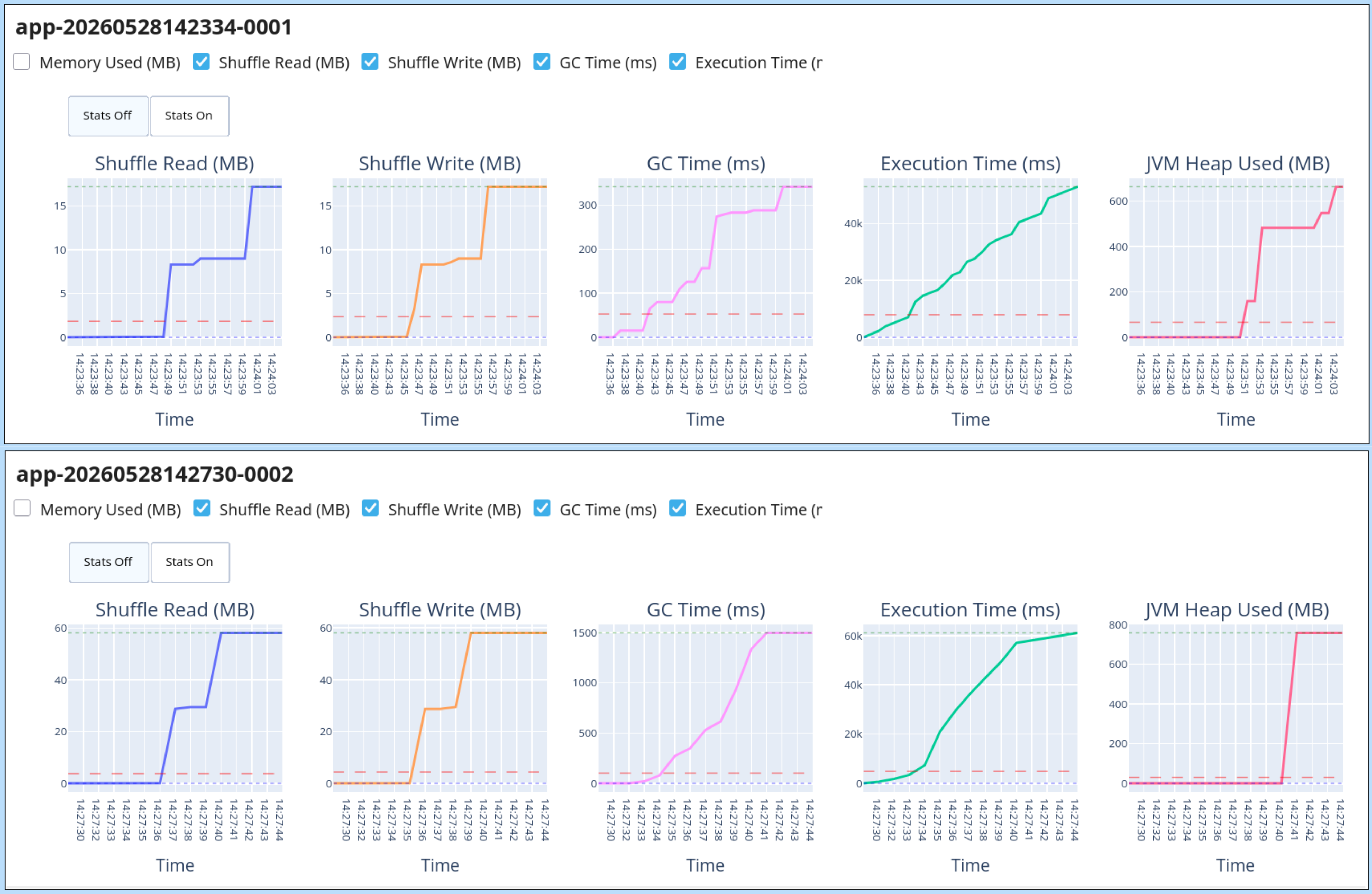}};
        \node[at=(main.north east), anchor=north east, shift=({-5mm,-3mm}), draw]{Phase I};
        \node[at=(main.east), anchor=north east, shift=({-5mm,-3mm}), draw]{Phase II};
    \end{tikzpicture}
    }
    \caption{Performance Metrics for Executor}
    \label{fig:test-plot}
\end{figure}

\figurename~\ref{fig:test-plot} depicts the \textquote{Performance Metric} section of Phase I and Phase II. It can be observed that Phase II completes the computation in less than half of the wall-clock time of Phase I (timings on $X$-axis of Execution Time plot), even though the execution times (i.e. cumulative CPU times) remain comparable across both phases. GC Time increases in Phase II, likely due to higher object-allocation rates associated with the increased core count. The memory pressure is comparable for both phases. Collectively, these finding indicate that increased parallelism is effective strategy for this use-case, as the reduction in wall-clock time outweighs the minor increase in garbage collection overhead.

This use-case of application optimization demonstrates \packagename{}'s capability to streamline the initial workflow for performance tuning by integrating cluster lifecycle management and performance metric observation directly within the Jupyter Notebook, thereby enhancing accessibility for beginner and experienced users alike.

However, despite its capabilities, \packagename{} has some limitations that can impact its efficiency in certain scenarios. Currently, it is best suited for relatively small applications, as continuous plot updates can affect the availability of the Python kernel. %
Furthermore, IFrame and cross-origin restrictions within IPyWidgets prevent direct embedding of the BDF native Web-UI in the \packagename{} interface. Currently, \packagename{} only supports Apache Spark and Apache Flink. While these limitations do not compromise functionality, they may slightly affect user experience.

\section{Conclusion}
This work introduced \packagename{}, a comprehensive yet intuitive tool that streamlines the setup, management, and monitoring of BDF on HPC systems using a fully automated mechanism, right inside the familiar JupyterHub's (Jupyter Notebook) graphical user interface.

\packagename{} eliminates the entry barrier for beginners and makes the process efficient for experts, encouraging broader adoption of data processing on HPC resources. By providing performance information, users can make informed decisions about resource allocation and code optimization, an important measure which is often overlooked by cross-domain users.
This capability is a concrete step towards a more accessible HPC environments, which is particularly important as data-processing demands surge with the growth of data intensive workloads.

Future work will focus on addressing the limitations of the current implementation, including improving the retrieval of additional data from the BDF Web-UI to provide more comprehensive insights. Additionally, optimizing widget performance will be a key area of focus to improve overall usability and responsiveness. Furthermore, developing an implementation that is independent of the current Python kernel is also planned, allowing for greater flexibility and compatibility, which will contribute to a more robust and user-friendly experience.

\begin{credits}
\subsubsection{\ackname}
The authors acknowledge the financial support by the Federal Ministry of Research, Technology and Space of Germany and by Sächsische Staatsministerium für Wissenschaft, Kultur und Tourismus in the programme Center of Excellence for AI-research \textquote{Center for Scalable Data Analytics and Artificial Intelligence Dresden/Leipzig}, project identification number: ScaDS.AI\\
The authors gratefully acknowledge the GWK support for funding this project by providing computing time through the Center for Information Services and HPC (ZIH) at TU Dresden.

\subsubsection{\discintname} The authors have no competing interests to declare that are relevant to the content of this article.
\end{credits}

%
%
\bibliographystyle{splncs04}
\bibliography{ref}
\end{document}